\documentclass[conference]{IEEEtran}
\usepackage{graphicx}
\usepackage{amssymb,amsmath,bm}
\usepackage{textcomp}

\usepackage{multirow}
\usepackage{epstopdf}
\usepackage{array}

\usepackage{cite}
\usepackage{url}

\newcolumntype{L}[1]{>{\raggedright\let\newline\\\arraybackslash\hspace{0pt}}m{#1}}
\newcolumntype{C}[1]{>{\centering\let\newline\\\arraybackslash\hspace{0pt}}m{#1}}
\newcolumntype{R}[1]{>{\raggedleft\let\newline\\\arraybackslash\hspace{0pt}}m{#1}}

\newcommand{\specialcell}[2][c]{%
  \begin{tabular}[#1]{@{}c@{}}#2\end{tabular}}

\ifCLASSINFOpdf
\else
\fi

\begin{document}
%
\title{Incorporation of Speech Duration Information in Score Fusion of Speaker Recognition Systems}

\author{\IEEEauthorblockN{Ali Khodabakhsh, Seyyed Saeed Sarfjoo,\\ Osman Soyyigit, Cenk Demiro\u{g}lu}
\IEEEauthorblockA{Electrical and Computer Engineering Department\\
\"{O}zye\u{g}in University, 34794, Cekmekoy, Istanbul, Turkey\\
Email: \{ali.khodabakhsh, saeed.sarfjoo, osman.soyyigit\}@ozu.edu.tr,\\cenk.demiroglu@ozyegin.edu.tr \vspace{-23pt}}
\and 
\IEEEauthorblockN{Umut Uludag}
\IEEEauthorblockA{TUBITAK BILGEM, 41470, Gebze, Kocaeli, Turkey\\
Email: umut.uludag@tubitak.gov.tr}
}


%


\maketitle

\begin{abstract}
In recent years identity-vector (i-vector) based speaker verification (SV) systems have become very successful.
Nevertheless, environmental noise and speech duration variability still have a significant effect on degrading the performance of these systems.
In many real-life applications, duration of recordings are very short; as a result, extracted i-vectors cannot reliably represent the attributes of the speaker.
Here, we investigate the effect of speech duration on the performance of three state-of-the-art speaker recognition systems.
In addition, using a variety of available score fusion methods, we investigate the effect of score fusion for those speaker verification techniques to benefit from the performance difference of different methods under different enrollment and test speech duration conditions.
This technique performed significantly better than the baseline score fusion methods.
\end{abstract}


%
\IEEEpeerreviewmaketitle

\vspace{-5pt}
\section{Introduction}
\label{sec:intro}

Over recent years, following the success of the identity vector (i-vector) based speaker verification (SV) methods \cite{dehak2011front}, these systems have made significant progress \cite{lee2013speaker}.
As speaker recognition technology reaches its maturity, real-life applications impose a drastic limitation for the systems in terms of environmental noise and speech duration variability during authentication.
The problem of duration variability is known to be one of importance for practical speaker recognition applications, and has also been addressed to a certain extent in the literature in the context of i-vector based speaker recognition systems \cite{garcia2013subspace}.
Furthermore, in the biannual speaker recognition evaluation (SRE) challenge held by National Institute of Standards and Technology (NIST), in year 2014, NIST coordinated a special i-vector challenge \cite{greenberg2014nist}, where the duration variability was one of the dominant challenges.

Most of the studies on i-vector based speaker recognition focus on recognition problems, where i-vectors are extracted from speech recordings of sufficient length.
Therefore, the majority of modeling techniques simply assume that the extracted i-vectors give a reliable estimation of the attributes of the speaker.
However, since the duration of recordings can be very short (on the order of less than 5 seconds) in many real-life applications, such as speaker identification tasks in broadcast data, this assumption fails to hold.

Only recently, a number of solutions have been proposed addressing the problem of duration variability.
For example, in \cite{garcia2013subspace}, \cite{kenny2013plda}, and \cite{cumani2013probabilistic}, authors do not treat i-vectors as point estimates of the hidden variables in the eigenvoice model, but rather as random vectors.
In this slightly different perspective, the i-vectors appear as posterior distributions, parameterized by the posterior mean and the posterior covariance matrix.
In \cite{borgstrom2013supervector} Borgstrom and McCree proposed a framework of super vector bayesian speaker comparison (SV-BSC), which keeps account of the observation noise throughout modeling and scoring.
In \cite{garcia2013subspace} with expanding the work in \cite{borgstrom2013supervector} Garcia-Romero and McCree reformulate SV-BSC and reach to SV-PLDA that facilitates the use of practical techniques, such as length normalization, and multi-cut enrollment averaging.
In \cite{vesnicer2014incorporating} Vesnicer et al. address the problem of duration variability through weighted statistics and demonstrate how established feature transformation techniques regularly used in the area of speaker recognition, such as PCA or WCCN, can be modified to take the duration into account.

In \cite{rastoceanu2011score}, Rastoceanu and Lazar make a comparison of different features and methods for score fusion for an independent speaker verification application.
In this paper, scores obtained with several types of features were fused with combination methods such as mean, max, min, weighted sum, and classification methods such as: support vector machines (SVM), linear discriminant analysis (LDA).
Based on the result of this paper, fusion methods outperformed the baseline GMM-UBM method.
In \cite{hautamaki2013sparse}, for fusion of different classifiers in speaker verification systems, Hautamaki et al. use classifier ensemble selection, which can be seen as sparse regularization applied to logistic regression.
However, none of the mentioned studies take duration variability into account for score fusion.

In this paper we investigate the effect of speech duration on the performance of three speaker recognition systems representing the current state-of-the-art: Gaussian Mixture Model-Universal Background Model (GMM-UBM) \cite{reynolds2000speaker}, Total Variability Space (TVS) \cite{dehak2011front,garcia2011analysis} and Probabilistic Linear Discriminant Analysis (PLDA) \cite{li2012probabilistic} scoring in TVS.
Furthermore, using a wide range of available score fusion methods, we investigate the effect of score fusion of these speaker verification techniques, to benefit from the performance difference of different methods under different enrollment and test speech duration conditions.

This paper is organized as follows. A background on speaker recognition, and description of speaker recognition systems and score fusion methods used in this study are given in Section \ref{sec:overview}.
Experimental setup and results are presented and discussed in Sections \ref{sec:experiments} and \ref{sec:results}, respectively. Finally, conclusion and future works are given in Section \ref{sec:conclusion}.

\section{Overview of speaker recognition systems}
\label{sec:overview}



\subsection{Feature extraction}

Feature extraction as transformation of the speech signal to a set of feature vectors representing the desired attribute, can be done using several different features.
In this study, similar to most studies involving speech and speaker recognition, the Mel-Frequency Cepstral Coefficients (MFCCs) are used due to their better performance compared to other features \cite{davis1980comparison}.

Sometimes to improve the robustness of features to channel differences, feature normalization methods such as cepstral mean subtraction \cite{westphal1997use}, cepstral mean and variance normalization, and feature warping are used.
In this study we used feature warping \cite{pelecanos2001feature}.

To increase the quality and effectiveness of modeling, non-speech frames need to be discarded prior to modeling.
Here, voice activity detection (VAD) is done using bi-gaussian modeling of speech frames on log energy distribution of the input frames.

\subsection{Speaker Modelling}

\subsubsection{GMM-UBM}

GMMs are typically used to represent the acoustic feature space in speaker recognition systems \cite{dehak2011low}.

In this method, first, using expectation-maximization (EM) algorithm \cite{reynolds1992integrated}, a GMM called universal background model (UBM) is trained from multiple sessions from multiple speakers.
For the enrollment step, given an utterance, the speaker model is adapted from UBM using maximum posterior adaptation (MAP) \cite{goronzy1999combined}.
Typically, only the mean is adapted.

\subsubsection{Total variability space (TVS)}

The supervector of mean vectors in UBM is of a very high dimensionality, and the number of parameters to adapt is very high.
Assuming speech consists of a speaker factor and an additive channel factor, speech model can be formulated as \vspace{-4pt}
\begin{equation}
M_s = S + C ,
\label{eq:FA}
\vspace{-2pt}
\end{equation}
where $M_s$ is speaker and channel dependent supervector, $S$ is speaker dependent supervector and $C$ is channel dependent supervector.
Eq. \ref{eq:FA} can be rewritten as \cite{dehak2011low}\vspace{-4pt}
\begin{equation}
M_s = M_0 + Vy + Ux + Dz ,
\label{eq:JFA}
\vspace{-2pt}
\end{equation}
where $M_0$ is speaker and channel independent mean supervector, $V$ is a low-rank eigenvoice matrix representing speaker space, $U$ is a low-rank eigenchannel matrix representing channel space, and $D$ is diagonal matrix which is modeling the Gaussian noise, and $x$, $y$, and $z$ are random vectors with a standard normal prior \cite{dehak2011low}. These vectors can be jointly computed using the joint factor analysis (JFA) approach.

In \cite{dehak2011front} Dehak et al. have shown that even though JFA is successful in increasing the performance of the recognition systems, there remains speaker variability in the channel factor.
They proposed a method combining both factors in a single matrix know as T matrix. \vspace{-4pt}
\begin{equation}
M_s = M_0 + Tw
\label{eq:TVS}
\vspace{-2pt}
\end{equation}
where $M_0$ is speaker and channel independent mean supervector, $T$ is a rectangular matrix of low rank, and $w$ is the identity vector (i-vector) and is a random variable with an standard posterior distribution.
In this approach, the speaker and channel factors are combined into a single vector $w$ in a lower dimensional space, postponing the speaker and channel factor separation task.

\subsection{Scoring}

For GMM-UBM system, given a set of feature vectors $X$ and a speaker model $M_{hyp}$, the similarity score is computed as \vspace{-4pt}
\begin{equation}
score = \log p(X|M_{hyp}) - \log p(X|M_0)
\label{eq:llr}
\vspace{-2pt}
\end{equation}
For TVS system, an i-vector is extracted from the test utterance, and compared to the i-vector extracted from the enrollment utterance using similarity measures.
In this study, similarity comparison is done using cosine distance scoring, and probabilistic linear discriminant analysis (PLDA) \cite{cumani2013probabilistic}. 
LDA and length normalization were used prior to PLDA in this study \cite{garcia2011analysis}.

One of the known methods that improve the accuracy of speaker recognition systems, and emphasized by the speaker recognition community, is score fusion of multiple subsystems.
Score fusion takes advantage of the fact that different systems make different mistakes, and by combining their output scores, the overall system can reduce the dependence of output decisions on the mistakes of a particular system.

In this study, we focus on simple mean, logistic regression (LR) and neural networks (NN) methods.

\section{Experiment setup}
\label{sec:experiments}

All systems in the experiments were trained with 19 dimensional MFCC features plus log-energy coefficient along with their delta and delta-delta parameters. 25 msec window with 10 msec window shift is used for feature extraction. Static log-energy feature is excluded, making the final dimensionality of features 59. Feature warping is done on each 300 frames (3sec windows). Bi-gaussian VAD is done using same windowing parameters as feature extraction. Details of training, development and test experiments are shown below.

\subsection{Training}

For training the speaker recognition system, VoxForge online corpus \cite{VoxForge} which is a user generated corpus is used.
This decision was motivated due to huge number of speakers and short-duration of the utterances in this corpus.
The corpus consists of many English dialects from native and non-native speakers.
Due to a big unbalance between the dialects and genders, we decided to limit the study to only male speakers with North American English dialect.
A summary of the training data is given in Table \ref{table:data}.

\begin{figure}
\vspace{-5pt}
\centering
\centerline{\includegraphics[width=0.48\textwidth,trim=80 300 90 300,clip]{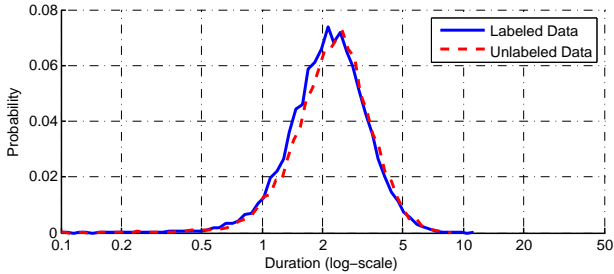}}
\vspace{-5pt}
\caption{Normalized histogram of distribution of training data (log-scale).}
\label{fig:distributions}
\vspace{-5pt}
\end{figure}

This user generated corpus consists of two parts, registered speakers and anonymous speakers.
Speech from registered users, were labeled based on their usernames and used for the supervised modeling steps (LDA and PLDA).
For models that did not require labels, the whole data was used. Furthermore, the maximum number of sessions per speaker was limited.
Distribution of these sets are shown in Fig. \ref{fig:distributions}.


A UBM consisting of 1024 gaussians was trained on all the available data. The same data was used for training of the T matrix with a rank of 500.
The ranks of LDA and PLDA models were set to 150 and 75 respectively, according to the best performance achieved on the training data.

\begin{table}
\centering
\caption{Databases, number of speakers, and number of sessions that were used for training, development, and test.}
\label{table:data}
\vspace{-5pt}
\begin{tabular}{l|c|c|c|c|}
\cline{2-5}
                                 & \multicolumn{2}{c|}{Training} & \multirow{2}{*}{Devel} & \multirow{2}{*}{Test} \\ \cline{2-3}
                                 & Labeled      & Unlabeled      &                        &                       \\ \hline
\multicolumn{1}{|l|}{Corpus    } & \multicolumn{2}{c|}{VoxForge} & WSJ0                   & WSJ1                  \\ \hline
\multicolumn{1}{|l|}{Sessions} & 19009        & 9637           & 822                & 2625              \\ \hline
\multicolumn{1}{|l|}{Speakers  } & 521          & -              & 66                     & 152                   \\ \hline
\end{tabular}
\vspace{-5pt}
\end{table}

\subsection{Development and test}

To minimize the effect of channel difference and to keep the focus on duration variability, Wall Street Journal (WSJ0, and WSJ1) corpus were chosen as development and test sets respectively.
To be able to analyze the effect of speech duration, a wide range of enrollment and test speech duration were targeted (0.1, 0.2, 0.5, 1, 2, 5, 10, 20, and 50 seconds).
The motivation behind the aforementioned duration setup was that typically segment duration distribution follows a log normal distribution.

To generate this data, speech data of each speaker from WSJ corpus was first concatenated, then split to sessions containing 50 seconds of speech.
For each speaker, one session is chosen for enrollment and maximum of 50 of the remaining sessions were used for test/development.
As we want to investigate the effect of different enrollment and test durations, we randomly generate 8 shorter sessions from each session with approximately 0.1 to 20 seconds of human speech. For applying a feature warping, minimum recording duration of each part should be more than 3 seconds. By doing this, 9 set of enrollment scenarios and 9 set of test/development scenarios are created.

Details about number of speakers and number of sessions for development and tests data can be found in Table \ref{table:data}.

For the sake of simplicity, only T-norm score normalization is done on the output of the subsystems \cite{auckenthaler2000score}. All identification tests are done under closed-set conditions.
Logistic regression (LR) score fusion is done using BOSARIS toolkit \cite{brummer2011bosaris}.
For neural network (NN) score fusion, a feed-forward network consisting 4 nodes in the hidden layer was used.

\section{Results and discussion}
\label{sec:results}

\begin{table*}[ht]
\centering
\caption{Speaker identification errors and speaker verification equal error rates (EER) are shown for all subsystems and score fusion methods. Results with duration-based score fusion are also reported where applicable.
Systems with best performance for identification and verification tests are shown as bold. For identification and verification experiments we have $2625\times81$ and $2625\times81\times152$ tests respectively. Distance between upper 95\% confidence interval and error mean for identification and verification tests are 0.1\% and 0.02\% respectively.  }
\label{table:whole_exp}
\vspace{-5pt}
\begin{tabular}{ll|r|r|r|c|c|c|}
\cline{3-8}
                                                                           &                & \multicolumn{3}{c|}{Subsystems}                             & \multicolumn{3}{c|}{Score fusion}       \\ \cline{3-8}
                                                                           &                & GMM   & \specialcell{TVS-cosine} & \specialcell{TVS-PLDA} & mean  & LR             & NN             \\ \hline
\multicolumn{1}{|l|}{\multirow{2}{*}{\specialcell{Identification Error}}}  & Overall        & 52.26 & 49.37                     & 46.42                   & 46.11 & \textbf{44.25} & 44.77          \\ \cline{2-8}
\multicolumn{1}{|l|}{}                                                     & Duration-Based & -     & -                         & -                       & -     & 44.95          & \textbf{44.36} \\ \hline
\multicolumn{1}{|l|}{\multirow{2}{*}{\specialcell{Verification EER}}}      & Overall        & 19.46 & 15.92                     & 15.89                   & 15.44 & 15.14          & 15.01          \\ \cline{2-8}
\multicolumn{1}{|l|}{}                                                     & Duration-Based & -     & -                         & -                       & -     & \textbf{13.39} & \textbf{13.46} \\ \hline
\end{tabular}
\vspace{-5pt}
\end{table*}

Two sets of experiments were done on the test data.
In the first set of experiments, all 81 sets of data were combined and accuracy of each subsystem as well as the results for score fusion using mean, LR, and NN were calculated.
Score fusion models were trained using the development data.
In the second experiment, LR and NN score fusion models were trained on each enrollment and development duration condition separately, creating 81 model for each system.
Accordingly, the tests were done using the models trained with the corresponding enrollment and test duration, and accuracies were reported on the whole data.
These results are shown in Table \ref{table:whole_exp}.
It is important to note that the high error rates in this table are caused by the very short tests durations (0.1, 0.2, 0.5, and 1 seconds).
In fact, experiments with very short durations on at least one side of enrollment or test cover 69\% of the total experiments.

One might expect that training a score fusion model for each duration condition will reduce the performance of the fused scores by limiting the amount of training data.
However, as it is seen on the Table \ref{table:whole_exp}, the best performance achieved for speaker verification system is achieved with duration-based system, reducing the error by 1.75\%.

\begin{figure*}

\begin{minipage}[b]{0.46\linewidth}
  \centering
  \centerline{\includegraphics[width=1\textwidth,trim=80 200 90 190,clip]{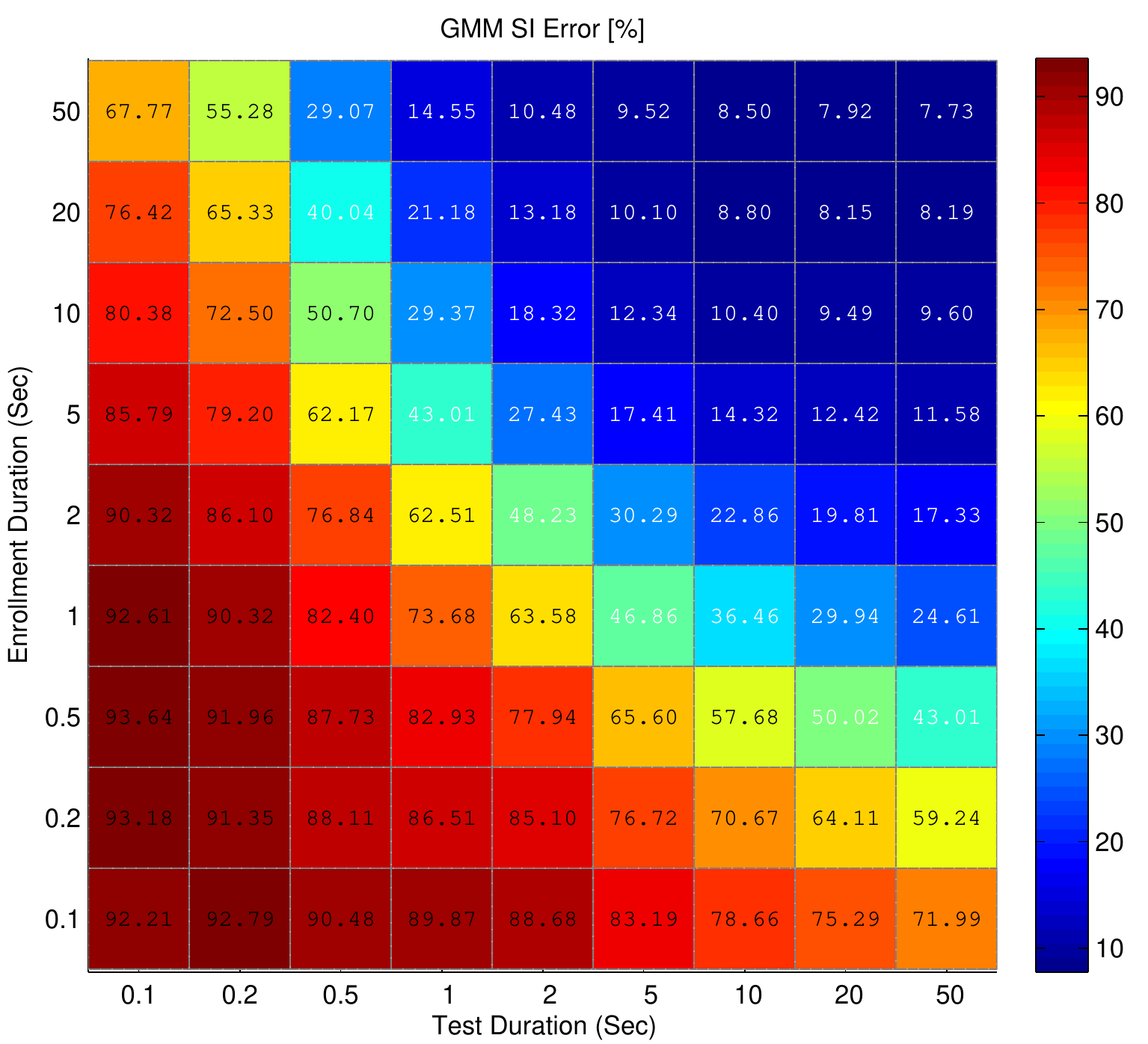}}
  \small\centerline{\specialcell{(a) GMM identification error rates for different \\ enrollment and test duration conditions }}\medskip
  \vspace{-2pt}
\end{minipage}
\hfill
\begin{minipage}[b]{0.46\linewidth}
  \centering
  \centerline{\includegraphics[width=1\textwidth,trim=80 200 90 190,clip]{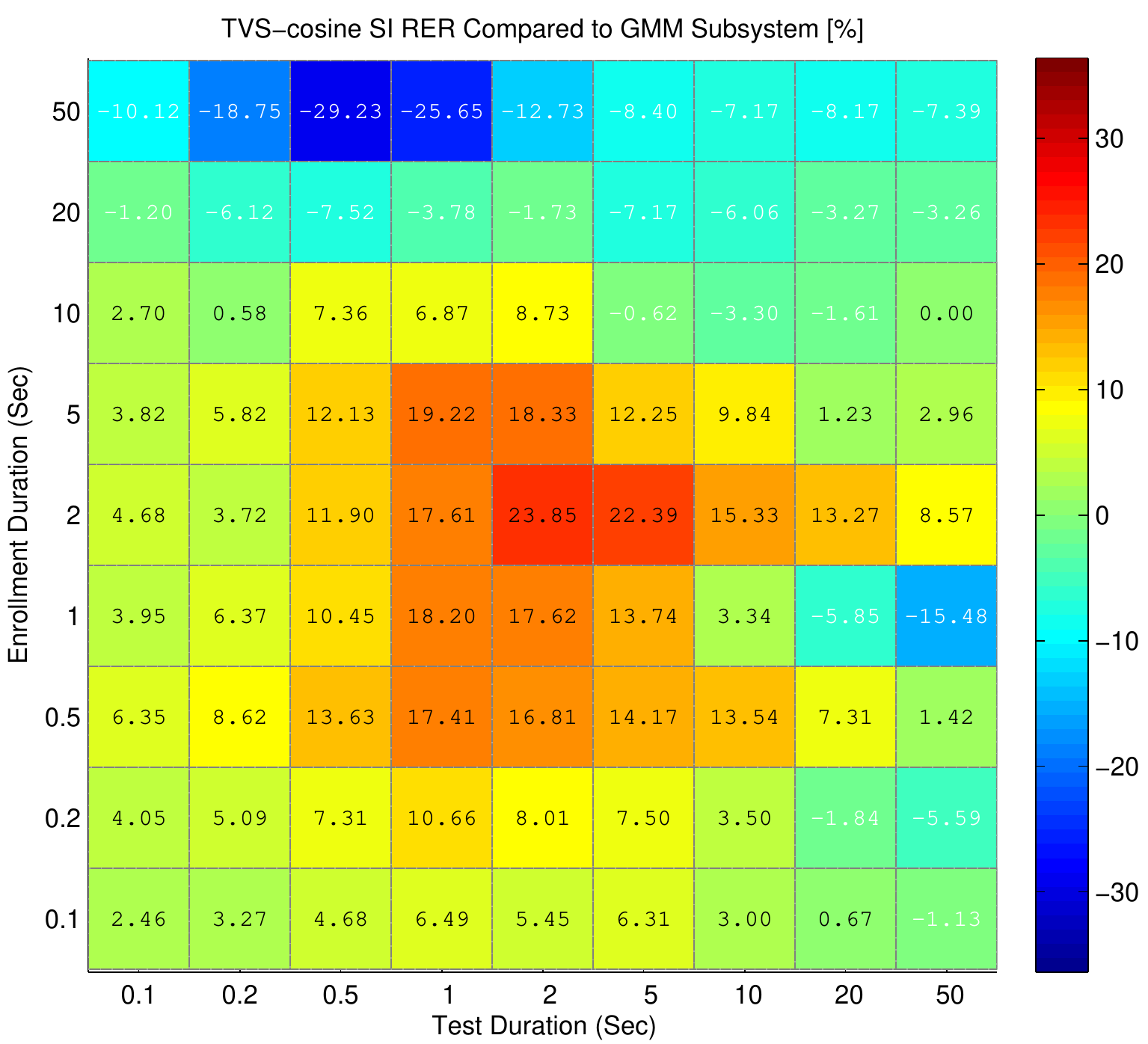}}
  \small\centerline{\specialcell{(b) Relative error reduction of TVS-cosine \\ compared to GMM system}}\medskip
  \vspace{-2pt}
\end{minipage}
\begin{minipage}[b]{0.46\linewidth}
  \centering
  \centerline{\includegraphics[width=1\textwidth,trim=80 200 90 190,clip]{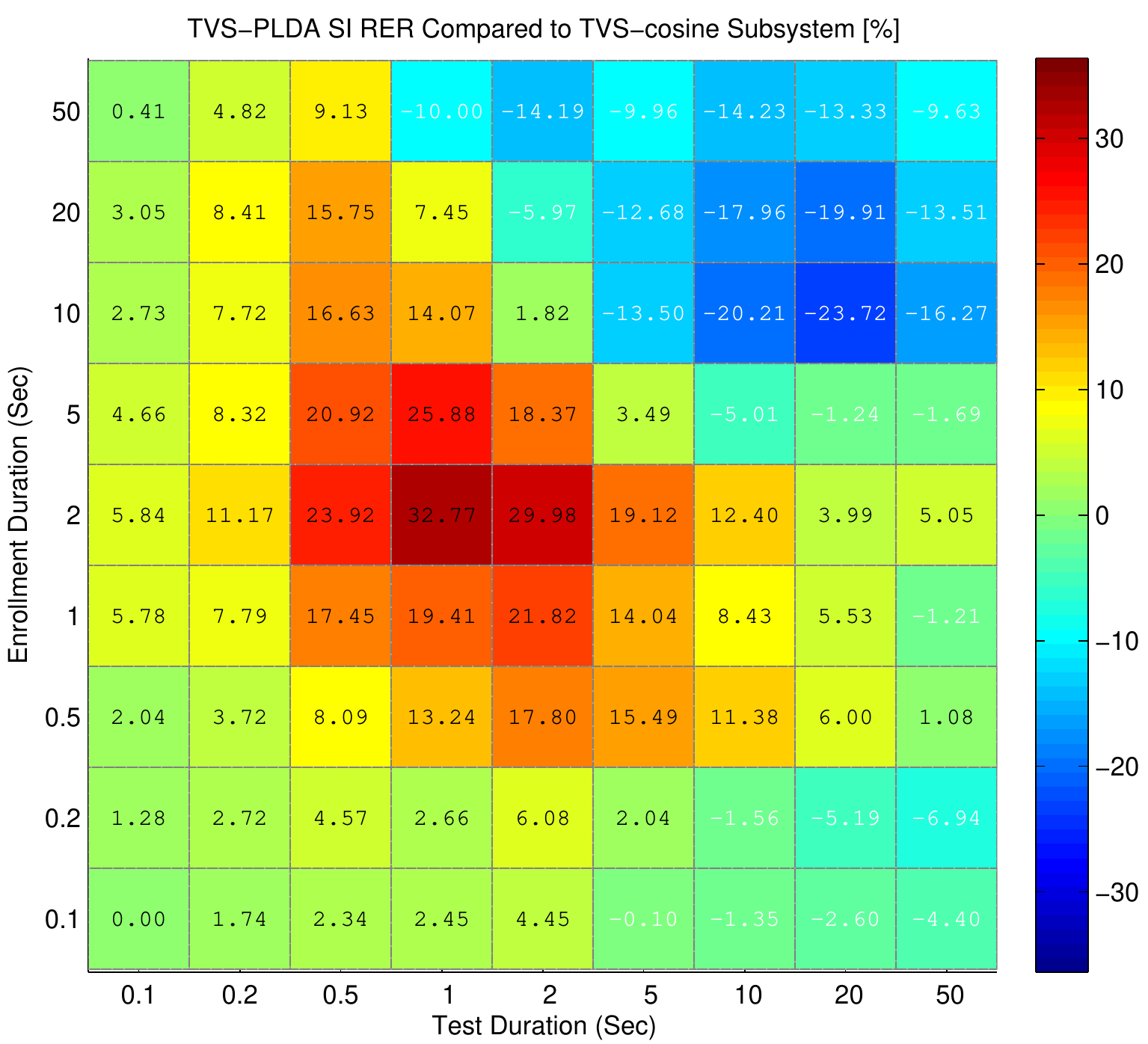}}
  \small\centerline{\specialcell{(c) Relative error reduction of TVS-PLDA \\ compared to TVS-cosine system}}\medskip
\end{minipage}
\hfill
\begin{minipage}[b]{0.46\linewidth}
  \centering
  \centerline{\includegraphics[width=1\textwidth,trim=80 200 90 190,clip]{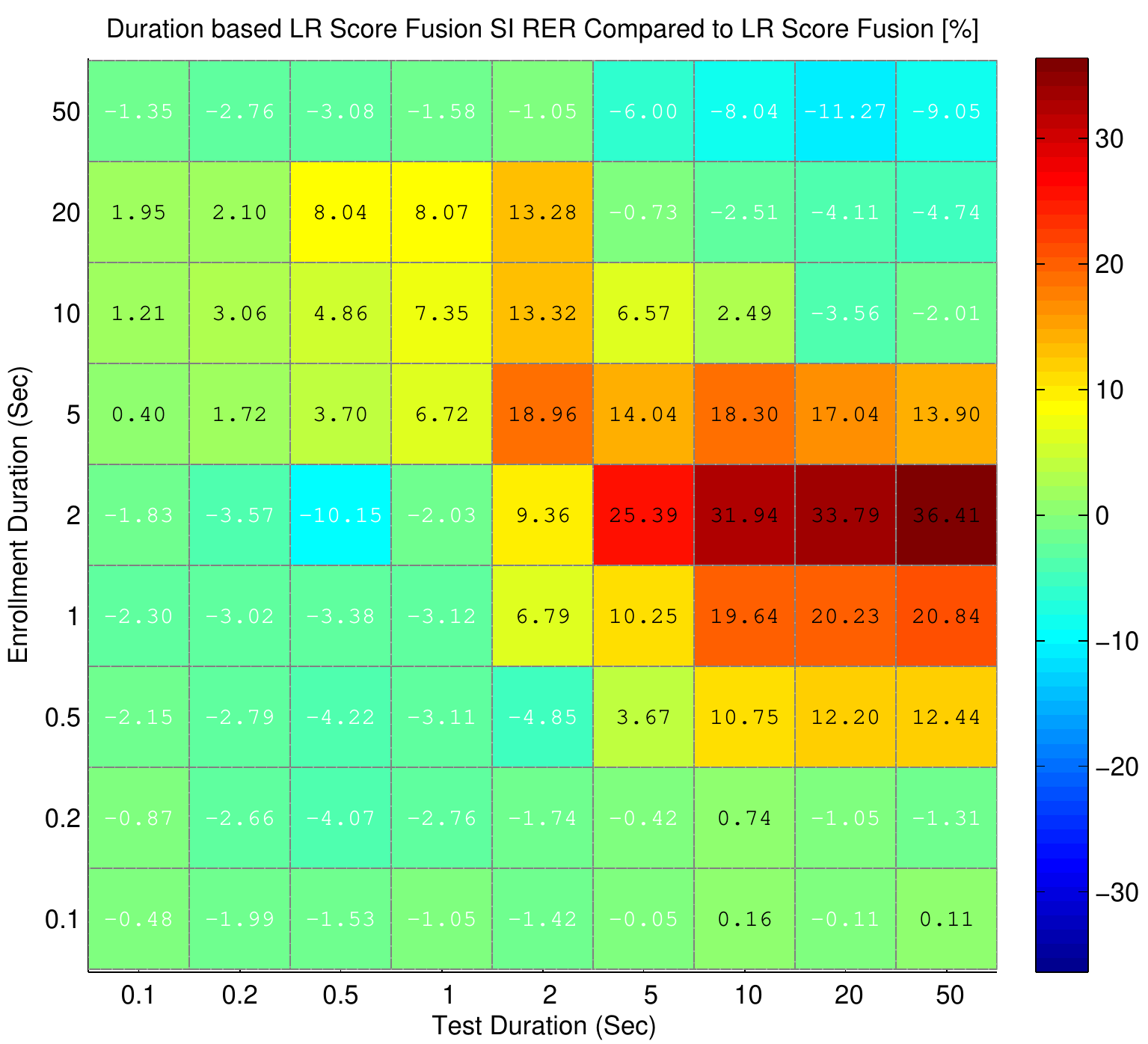}}
  \small\centerline{\specialcell{(d) Relative error reduction of Duration-based LR \\ compared to LR Score Fusion}}\medskip
\end{minipage}
\vspace{-12pt}
\caption{Error of the GMM subsystem (a), along with the relative error reduction (RER) for speaker identification (SI) for special conditions.}
\label{fig:errors}
\vspace{-14pt}
\end{figure*}

To gain insight on the reason for this behavior, a third set of experiments are done on each duration condition, and results were reported independently.
In Fig. \ref{fig:errors}a the error rates of the GMM method is shown.
As expected, as the duration of test and enrollment increases, the error rate becomes lower.
Due to space saving, only the results for speaker identification experiments are visualized.

To observe the effectiveness of TVS-cosine method, the relative error reduction (RER) rate of this system compared to the base GMM system is shown in Fig. \ref{fig:errors}b.
It can be seen that the biggest relative increase of accuracy occurred around 2 second enrollment and test, which correspond to the training distribution shown in Fig. \ref{fig:distributions}.
On the other hand, there is a stable increase of accuracy where the length of enrollment and test is very short, which corresponds to better modeling of TVS-cosine for these situations due to reduction of number of parameters for adaptation as mentioned before.
Another interesting observation made is that the TVS-cosine fails to increase accuracy in most cases when the length of enrollment is longer than 20 seconds.

The relative error reduction of TVS-PLDA compared to TVS-cosine is visualized in Fig. \ref{fig:errors}c.
Here a similar effect to previous case can be observed.
These results show the dependence of the accuracy gain of PLDA to the duration of enrollment and test, as well as the distribution of training data.

To see how the duration-based LR takes advantage of the dependency of performance of different systems to durations, the relative error reduction of duration-based LR compared to LR is visualized in Fig. \ref{fig:errors}d.
The normalized weights of 0.17, 0.37, and 0.46 were assigned to GMM, TVS-cosine, and TVS-PLDA respectively by the LR model.

In this figure, it is observable that in most of the cells there is a small increase in the error rate.
This event was expected and can be explained by significant reduction in amount of training data used for training the duration-based LR compared to overall LR.
On the other hand, where the enrollment duration matches the distribution of training data (Fig. \ref{fig:distributions}) and the duration of test is longer than 2 seconds, it can be seen that the duration-based LR could reduce the error by up to 66\% ratio.

It also can be seen that even though by limiting the score fusion training data for each case we expect a lower accuracy gain in the duration based score fusion, the extra accuracy gained comes from special conditions where the performance of the subsystems vary a lot for different duration of enrollment and test.

\section{Conclusion and future work}
\label{sec:conclusion}

In this study we investigated the effect of duration of enrollment and test sets on different state-of-the-art speaker recognition systems. Furthermore, using several score fusion methods, we investigated the effect of score fusion
of these speaker verification techniques, to benefit from the
performance difference of different methods under different
enrollment and test speech duration conditions. Based on our observations, duration-based technique performed better than the baseline overall score fusion methods. When we compared the accuracy of duration-based methods with overall baseline methods, it was observed that there is a dependency between the gained accuracy of TVS-based  methods, and duration of enrollment and test sets, as well as the distribution of the data used for training the T matrix and gained accuracy of PLDA model are correlated.
It was also observed that as the duration of enrollment and test recordings increases, the TVS-cosine and TVS-PLDA methods, with respect to GMM-UBM method give a lower gain in accuracy.

These observations motivates us to investigate the possibility of taking advantage of this performance difference of different systems for a more effective score fusion method.
To this goal we investigated training a separate score fusion model for each duration condition, however this method could not give the expected gain in accuracy due to the fact that by splitting the available data for training score fusion, and training multiple models, the effectiveness of these models decrease.
However, still some improvement in the overall accuracy of the systems was observed.

To prevent this condition, based on the dependence of the accuracy gain of TVS-cosine and TVS-PLDA methods to the duration of their training data, we hypothesized that there can be found a function for predicting the duration dependent weights of each subsystem\vspace{-1pt}
\begin{eqnarray}
s = \sum\nolimits_{subsystems} w_is_i \\
w_i = f_i(d_{enroll},d_{test}|\Gamma_{train})
\label{eq:fusion}
\vspace{-7pt}
\end{eqnarray}
where $s$ is the final score, $s_i$ is the output score of the $i$th subsystem, $w_i$ is the corresponding weight of the $i$th subsystem, $f_i$ is a subsystem specific function, mapping the duration of enrollment $d_{enroll}$ and duration of test $d_{test}$ and distribution of the training data $\Gamma_{train}$ to the desired weight.

Further studies are needed to formulate such functions for each subsystem.

\bibliographystyle{IEEEtran}
\bibliography{refs}

\end{document}